%% file: main.tex
\documentclass[sigconf]{acmart}
\renewcommand\footnotetextcopyrightpermission[1]{} 
\def\etal{\emph{et al.}}

\usepackage[acronym]{glossaries}
\usepackage{acronym}
\newacronym{ml}{ML}{machine learning}
\newacronym{fr}{FR}{facial recognition}
\newacronym{fv}{FV}{facial verification}

\newacronym{pdf}{PDF}{probability density function}

\newacronym{cnn}{CNN}{convolutional neural network}
\newacronym{nn}{NN}{neural network}
\newacronym{mtcnn}{MTCNN}{\emph{multi-task \gls{cnn}}}

\newacronym{gan}{GAN}{generative adversarial network}
\newacronym{se}{SE}{\emph{Squeeze-and-Excitation}}
\newacronym{d}{$D$}{discriminator}
\newacronym{g}{$G$}{generator}
\newacronym{ae}{AE}{autoencoder}
\newacronym{vae}{VAE}{variational autoencoder}
\newacronym{dbvae}{DB-VAE}{Debiasing Variational Autoencoder}

\newacronym{lut}{LUT}{Look-Up-Table}
\newacronym{soa}{SOTA}{state-of-the-art}

\newacronym{fiw}{FIW}{Families In the Wild}
\newacronym{lfw}{LFW}{Labeled Faces in the Wild}
\newacronym{bfw}{BFW}{Balanced Faces In the Wild}
\newacronym{rfw}{RFW}{Racial Faces in-the-Wild:}
\newacronym{dp}{DemogPairs}{Demographic Pairs}
\newacronym{itwcc}{ITWCC}{Wild Child Celebrity}

\newacronym{fd}{FD}{Face Discrimination}
\newacronym{bb}{BB}{bounding box}

\newacronym{sdm}{SDM}{signal detection model}
\newacronym{roc}{ROC}{receiver operating characteristic}
\newacronym{nmse}{NMSE}{Normalized Mean Square Error}
\newacronym{det}{DET}{Detection error trade-off}
\newacronym{tp}{TP}{true-positive}
\newacronym{fp}{FP}{false-positive}

\newacronym{cs}{CS}{Cosine Similarity}
\newacronym{sota}{SOTA}{state-of-the-art}

\newacronym{lime}{LIME}{Local Interpretable Model-Agnostic Explanations}
\newacronym{nas}{NAS}{Neural Architecture Search}
\newacronym{gapf}{GAPF}{Generative Adversarial Privacy and Fairness}
\newacronym{rnn}{RNN}{Recurrent Neural Network}
\newacronym{tclstm}{TC-LSTM}{Temporal Convolution-LSTM}

\newacronym{tcn}{TCN}{Temporal Convolution Network}

\newacronym{lstm}{LSTM}{long short-term memory}
\newacronym{hmm}{HMM}{Hidden Markov Models}
\newacronym{gmm}{GMM}{Gaussian Mixture Model}
\AtBeginDocument{%
  \providecommand\BibTeX{{%
    \normalfont B\kern-0.5em{\scshape i\kern-0.25em b}\kern-0.8em\TeX}}}

\acmConference[Zhuang et al.]{MM '20: ACM Multimedia}{2020}{TC-LSTM}

\begin{document}

\title{Towards 3D Dance Motion Synthesis and Control}


\author{Wenlin Zhuang}
\email{wlzhuang@seu.edu.cn}
\affiliation{%
  \institution{Southeast University}
}

\author{Yangang Wang}
\email{yangangwang@seu.edu.cn}
\affiliation{%
  \institution{Southeast University}
}

\author{Joseph Robinson}
\email{robinson.jo@husky.neu.edu}
\affiliation{%
  \institution{Northeastern University}
}

\author{Congyi Wang}
\email{artwang007@gmail.com}
\affiliation{%
  \institution{XMov}
}

\author{Ming Shao}
\email{mshao@umassd.edu}
\affiliation{%
  \institution{University of Massachusetts Dartmouth}
}

\author{Yun Fu}
\email{yunfu@ece.neu.edu}
\affiliation{%
  \institution{Northeastern University}
}

\author{Siyu Xia}
\email{xia081@gmail.com}
\affiliation{%
  \institution{Southeast University}
}








\renewcommand{\shortauthors}{Zhuang et al.}

\newcommand{\ie}{\textit{i}.\textit{e}., }
\newcommand{\eg}{\textit{e}.\textit{g}., }
\newcommand*{\etc}{etc.\@\xspace}

\newcommand{\xmark}{\ding{56}}%
\newcommand{\checkc}{\ding{51}}%
\newcommand{\NA}{---}

\begin{abstract}
3D human dance motion is a cooperative and elegant social movement. Unlike regular simple locomotion, it is challenging to synthesize artistic dance motions due to the irregularity, kinematic complexity and diversity. It requires the synthesized dance is realistic, diverse and controllable. In this paper, we propose a novel generative motion model based on temporal convolution and LSTM,~\gls{tclstm}, to synthesize realistic and diverse dance motion. We introduce a unique control signal, the~{\emph{dance melody line}}, to heighten controllability. Hence,  our model, and its switch for control signals, promote a variety of applications: random dance synthesis, music-to-dance, user control, and more. Our experiments demonstrate that our model can synthesize artistic dance motion in various dance types. Compared with existing methods, our method achieved start-of-the-art results.

\end{abstract}



\begin{teaserfigure}
  \includegraphics[width=\textwidth]{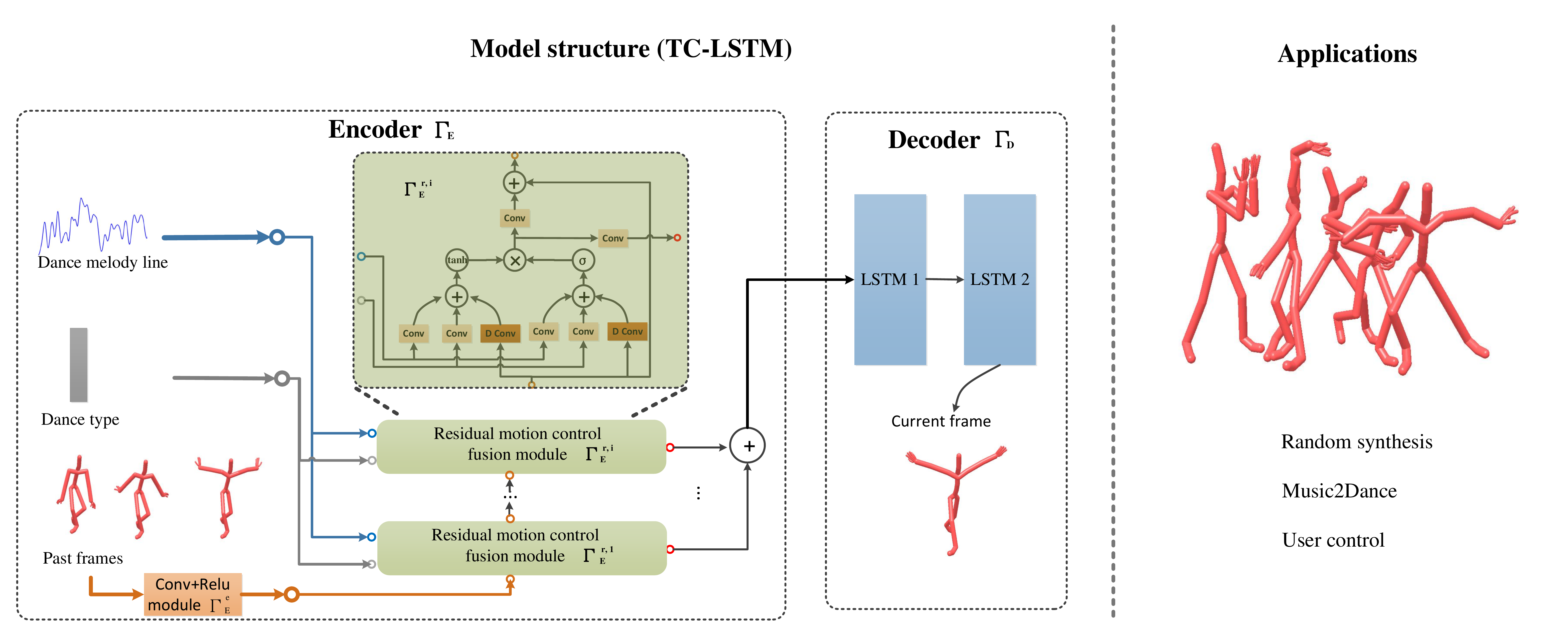}
  \caption{Framework of our TC-LSTM model. We take 3D dance synthesis as an autoregressive process conditioned on control signals. Our model can synthesize diverse and controllable dance motion for different dance types, and achieve multiple applications: random synthesis, Music2Dance and user control. "Conv": convolution, "D conv": dilated convolution.}
  \Description{Our framework.}
  \label{fig:teaser}
\end{teaserfigure}

\maketitle

\glsresetall
\input{introduction.tex}

\input{related_work.tex}

\input{overview.tex}

\input{data_processing.tex}

\input{model.tex}

\input{experiment.tex}

\input{conclusion.tex}


\bibliographystyle{ACM-Reference-Format}
\bibliography{sample-base}



\end{document}

%% file: introduction.tex
\section{Introduction}\label{sec:introduction}
Dance and concepts thereof are embroidered in our society, culture, and history~\cite{boyd2004dance}-- whether it is freestyle (\ie on-the-fly), a specific dance to a certain song (\eg the \emph{macarena}),  a spiritually or culturally inspired dance-off, or even just solo acts of dancing while alone and replaying a melody from memory. Hence, dance moves have the power to allow one-to-many individuals express emotions, all the while having the persistence to inspire, spread knowledge, show culture, and promote believes. As a part of the mobile-age, dance performances are easily main-streamed, making it possible to broadcast or view from just about anywhere at given moment. Now, we aim to best leverage our scientific research to directly enhance the melody of life we call dancing. 

Dancing can be considered a form of art. It requires professional choreographers to create and design artistic movements to express emotions. For this, professional dancers are trained and equipped with a rich repertoire of dance steps - the more creative, the better. Different dancers perform quite differently, even to the same music or melody. Nonprofessionals would typically find it challenging to create a dance. Therefore, to acquire the ability to automatically create dances is a daunting task, as dancing contains high kinematic complexity that span long-term spatio-temporal structures (\ie temporal 3D human dance motion), which make it difficult to synthesize realistic dance. 
More importantly, dance motion is diverse, irregular, complex, and often designed for specific music or melody. In addition, it is important to note that dancing is inherently a multi-modal problem~\cite{lee2019dancing}, spanning multiple views (\ie various dances for the same song). Lastly, different music or melodies should yield a whole variety of dance types. The challenges and specifications mentioned here demand an effective generative model to handle the complex and diverse dance motions. Furthermore, with such high powerful model, it should be adequate for various applications: freestyle(random synthesis~\cite{li2017auto}), dancing with music(music2dance~\cite{zhuang2020music2dance}), and support ordinary users to create dance(user control).

Early on, researches mainly adopted similarity-based retrieval methods to synthesize simple locomotion~\cite{li2002motion,min2012motion,safonova2007construction}. Then, others proposed methods to synthesize long-term dance motions in accordance to musical inputs~\cite{fan2011example,ofli2011learn2dance,lee2013music}.
However, these strategies lack flexibility, creativity, and are difficult to apply to irregular, complex dance motion. More recently, deep learning-based motion synthesis algorithms have shown higher potential~\cite{fragkiadaki2015recurrent, martinez2017human, li2017auto, peng2017deeploco, peng2018deepmimic}. The deep \gls{nn} can well model spatio-temporal structures with high kinematic complexity without taking up too much memory. \gls{rnn}~\cite{fragkiadaki2015recurrent, martinez2017human} has been proposed in recent years to model human motions for human motion prediction. However, these methods can easily fall into the temporal accumulation error(\ie get stuck in static poses). Li~\etal~proposed the auto-condition training strategy to train the model based on \glspl{rnn} to synthesize complex dance motion~\cite{li2017auto}. The method can only achieve random motion synthesis(long-term motion prediction), but not controllable motion synthesis. Considering the control signal, the method based on \glspl{lstm}~\cite{lee2018interactive} can synthesize controllable simple locomotion to interact with the environment, and it is difficult to synthesize complex controllable dance motion. Recently, some learning-based methods \cite{tang2018dance,lee2019dancing,zhuang2020music2dance} have been used to synthesize controllable dance motion from music. Lee~\etal~designed a \gls{nn} based on \gls{vae} and \gls{gan} to synthesize 2D dance movement~\cite{lee2019dancing}. The \glspl{lstm}-\gls{ae}~\cite{tang2018dance} is proposed to synthesize 3D dance motion from music feature, but the synthesized dance motion is far from realistic and diverse.The DanceNet, based on \gls{tcn}, have been applied to synthesize diverse dances for different dance types~\cite{zhuang2020music2dance}. However, the method can not synthesize various dances for the same music. Different from the direction and speed control in the locomotion synthesis~\cite{lee2018interactive}, the control signal adopted by the method is directly extracted from the music. Although there is the rhythmic consistency between music and dance, it is difficult to completely determine the dance motion. This is a weak (\ie not strong) control signal, so the controllability of their synthesized dance motion is lacking. Moreover, existing methods do not span various applications with the same model.


We first propose a dance control signal. Unlike in~\cite{zhuang2020music2dance}, we introduce a control signal, called \emph{dance melody line}, and it is highly correlated to the dance since it is extracted directly from the dance motion. We sum up the speed of the salient joints frame-by-frame to capture the melody control signal (\ie a 1D control signal). Provided a low-dimensional signal produced directly by motion, its coupling is notably strong, \ie different dance motions correspond to the same \emph{dance melody line}. This helps to synthesize different dance motions conditioned with the same control signal.

We also introduce a novel generative model with encoding and decoding stages. \glspl{tcn} are robust to noisy inputs~\cite{zhuang2020music2dance}, so we adopt it to extract motion features and fuse control signals to obtain controllable motion features, which is done as part of the the encoding stage. To strengthen the long-term spatio-temporal dependence of the output frames, we adopt the \glspl{lstm} as the decoder. Our model overcomes the shortcomings of the \glspl{lstm} that is not robust to noise, while ensuring that the output frames leverage long sequence dependency. Our model obtains the controllable motion features based on the \gls{tcn}, and then \glspl{lstm} decode to synthesize controllable dance motion. As a whole, we call our framework the \glspl{tclstm}. The output of \glspl{tclstm} is designed as a \gls{pdf} (\ie Gaussian mixture model), which also makes our model more robust.
With a careful training strategy (\ie mix training), our model supports switch melody control signal to synthesize dance motion, meaning that our model can use the same parameters for random and controllable motion synthesis.


\setlength{\tabcolsep}{4pt}
\begin{table*}[t]
\begin{center}
\caption{Compare against \gls{sota} methods about 3D dance synthesis for different applications. Model extensibility means the model can synthesis dance for different dance types. $\checkmark$ means the method can achieve the application. $\times$ means the method can not achieve the application.
}
\label{table:app_compare}
\begin{tabular}{l|c|c|c|c}
\hline
Method & Model extensibility & Random synthesis  & Music2Dance & User control \\
\hline
ac-\gls{lstm}\cite{li2017auto} & $\times$  & $\checkmark$  & $\times$  & $\times$  \\
\gls{lstm}-\gls{ae}\cite{tang2018dance}  & $\times$ & $\times$  & $\checkmark$  & $\times$  \\
DanceNet\cite{zhuang2020music2dance} & $\checkmark$ & $\times$  & $\checkmark$  & $\times$  \\
\hline
Ours & $\checkmark$ & $\checkmark$  & $\checkmark$  & $\checkmark$  \\

\hline	
\end{tabular}
\end{center}
\end{table*}
\setlength{\tabcolsep}{1.4pt}

 \vspace{2mm}
\textbf{Applications.}
We ran experiments on music-dance pair dataset\\~\cite{zhuang2020music2dance}. The results show that our model can generate realistic and diverse dance motions for different applications. Listed as follows:

\vspace{1mm}
\noindent\emph{Random synthesis.} We can switch off the melody control signal, and our model synthesizes long-term (\ie arbitrary length), diverse dance sequences(without the temporal accumulation error).

\vspace{1mm}
\noindent\emph{Music2Dance.} From the analysis of the music-dance paired data~\cite{zhuang2020music2dance}, we found that the melody lines of music and dance are highly matched, so we directly use the music melody line as the melody control signal. 
The type control signal can be obtained by a classifier as in~\cite{zhuang2020music2dance} or user given. Through the music melody line and dance type to synthesize the dance motion consistent with the melody and style of music.

\vspace{1mm}
\noindent\emph{User control.} The dance melody line is a 1D signal, which is easily given by ordinary users. Therefore, our approach allows ordinary users to design dance motions. We can synthesize the controllable dance through the user-defined melody line(i.e., drawing)  as the melody control signal.

\vspace{2mm}
\textbf{Research contributions.}
Along with the direct, tangible benefits, we propose the following contributions in research:

\vspace{1mm}
\noindent\emph{Controllability.} To our knowledge, we are the first to propose a controllable dance synthesis framework.

\vspace{1mm}
\noindent\emph{Robust to noisy inputs and long-term dependencies.} Our encoder-decoder structure ensures robustness to noise,and building long-term spatio-temporal dependence of the output frames.

\vspace{1mm}
\noindent\emph{\gls{sota} results with various applications.} Our experiments show that our approach can achieve \gls{sota} results for different applications.

%% file: related_work.tex
\section{Background}\label{sec:background}
We review three research area most related to the proposed (\ie motion synthesis and control, dance motion, and generative model).

\vspace{2mm}
\noindent\textbf{Motion synthesis and control}. 
Researchers tend to synthesize motions via data-driven methods, \ie, \gls{hmm}~\cite{bowden2000learning, brand2000style}, spatial-temporal dynamic models \cite{chai2007constraint, wei2011physically, lau2009modeling,xia2015realtime}, and low-dimensional statistical models \cite{chai2005performance, grochow2004style}. In addition, other methods to synthesize locomotion were based on motion graphs~\cite{li2002motion, lee2002interactive, kovar2008motion, min2012motion, safonova2007construction}. A common strategy among the aforementioned methods is the formulation of similarity-based retrieval to synthesize simple locomotion, which are completely dependent on the availability of dataset, hence, lacks flexibility and tend not to generalize well. Nowadays, deep \gls{nn}-based methods~\cite{fragkiadaki2015recurrent,holden2016deep} gradually started being used to synthesize motion. For instance, the \gls{rnn}-based methods that were proposed to predict short-term human motion, while being unable to synthesize long-term motion due to the temporal accumulation error~\cite{fragkiadaki2015recurrent,jain2016structural, martinez2017human}. Li~\etal~adopted the auto-conditioned training strategy to synthesize long-term motion, but lacking the ability to control the generated motion~\cite{li2017auto}. Phase-functional networks~\cite{holden2017phase} and \glspl{lstm}-based method~\cite{lee2018interactive} were introduced to synthesize controllable locomotion. However, the these methods are still limited - either by simple-random or simple-controlled locomotion, thus they are unable to synthesize complex dance motion with complete control. These limitations can be overcome in our approach. Our method can synthesize realistic, complex, diverse and controllable dance motion sequences.

\vspace{2mm}
\noindent\textbf{Dance motion}. As mentioned, earlier research tended to focus on synthesizing dance motions by adopting similarity retrieval strategies (\eg motion graph~\cite{li2002motion, lee2002interactive}). Fan~\etal~divided the long-term dance motion into multiple short-term clips, which were then used to build a motion graph~\cite{fan2011example}. Shiratori~\etal~retrieved each dance segment where the music and dance rhythm were consistent~\cite{shiratori2006dancing}. However, these methods rely entirely on the dataset, and lack the ability to truly creativity are music-consistency. Recently, various types of \glspl{nn} emerged as solutions to generate dance movements. For instance, \gls{vae} and \gls{gan} models were proposed to synthesize 2D dance motion from music~\cite{lee2019dancing}. Tang~\etal~built an \gls{lstm}-\gls{ae} to generate 3D dance motions; however, the generated motions are unrealistic~\cite{tang2018dance}. Li~\etal~proposed auto-conditioned \gls{lstm} to synthesize 3D dance motion; however, this work lacked motion-control(just random synthesis)~\cite{li2017auto}. A model based on temporal convolution was proposed to generate 3D, controllable dance motions, but the controllability is limited in its inability to synthesize multimodal dances provided the same control signals~\cite{zhuang2020music2dance}. Our model overcomes the limitations in being controllable, as we are able to synthesize realistic, diverse, and controllable multimodal dances. Furthermore, we use just a single model in a wide-range of applications: random synthesis, music2dance, and user control (Fig~\ref{fig:teaser}).

\vspace{2mm}
\noindent\textbf{Generative model}. In the motion synthesis model, the common autoregressive model is an \glspl{lstm}~\cite{fragkiadaki2015recurrent,jain2016structural, martinez2017human,li2017auto}. Like most others, these models can only generate random outputs, and lack the ability to control the synthesized motion. Lee~\etal~designed the control signal at the model input, but lacked robustness to noise~\cite{lee2018interactive}. Zhuang~\etal~proposed an autoregressive model based on temporal convolution~\cite{zhuang2020music2dance}, although the model is robust to input noise, it was unable to capture temporal dependencies across output frames. We carefully consider the pros and cons of the temporal convolution and \glspl{lstm}-based methods - leveraging the strengths of each to design our model. During the encoding phase, we train our model to be insensitive to input noise via temporal convolution to encode features. Thereafter, we employ an \glspl{lstm} to decode, as it enhances the temporal correlation of the output motion so that our method synthesizes realistic, complex, diverse and controllable dance motions.

%% file: overview.tex
\section{overview}\label{sec:overview}

The proposed framework is shown in Figure \ref{fig:teaser}.  We take 3D dance synthesis as an autoregressive process conditioned on control signals. specifically, to synthesize current frame, we take the previous dance frames and control signals(dance melody line and dance type) as inputs. We will introduce dance motion and control signal processing in Section \ref{sec:data_processing}. The model consists of two parts: encoder based on temporal convolution and decoder based on LSTMs. We will elaborate on our model in Section \ref{sec:model}. After decoding, the model outputs the \gls{pdf} of current dance frame, then we sample from the \gls{pdf} to get the current frame. Our model can realize different applications with a set of parameters: dance random synthesis, Music2Dance, and user control, which are introduced in Section \ref{sec:Experiment}.

%% file: data_processing.tex
\section{Data Processing}\label{sec:data_processing}
Zhuang~\etal~introduced a high-quality music-dance pair dataset to synthesize dance-from-music~\cite{zhuang2020music2dance}. This dataset consists of two types of dances - modern dance ($\approx$26.15 minutes, 94,155 frames at 60 FPS) and Korean dance ($\approx$31.72 minutes, 114,192 frames at 60 FPS) - we used to train the proposed model. The aim of this work was to generate controllable dance motions with the control signal made-up of two parts, \ie dance melody line (characterize the dance rhythm, local condition), and dance type(characterize the dance style, global condition). The dance type can be represented as an one-hot vector $c^{s}$, similar to ~\cite{zhuang2020music2dance}.
We next describe the dance melody line (Section~\ref{sec:dance_melody_line}), and then the dance motion representation (Section\ref{sec:motion_representation}).

\begin{figure}
	\centering
	\includegraphics[width=\linewidth]{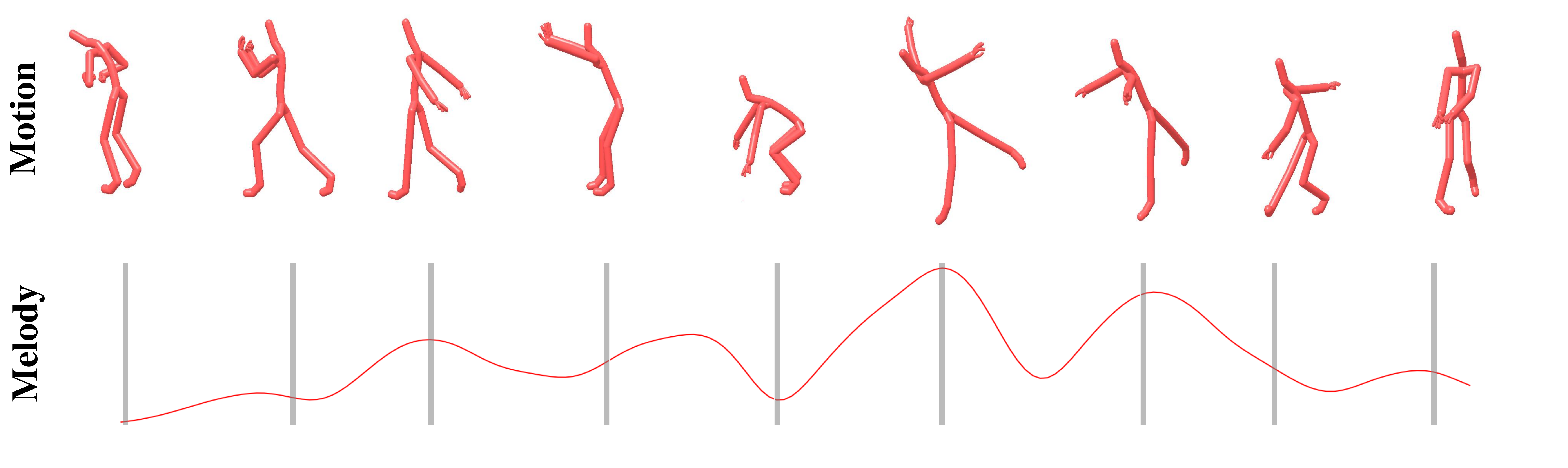}
	\caption{Depiction of the melody line from a modern dance. We proposed quantifying the melody lines as 1D signals.}
	\label{fig:dance_melody_line}
\end{figure}

\subsection{Dance melody line}\label{sec:dance_melody_line}
Professional choreographers taught us that human dance moves are more than just random movements~\cite{hewitt2005social}. There is an internal melody for dance founded on higher-level information that reflects the rhythm and speed of a dance, which mirrors that of the main melody line in music theory~\cite{mason2012music}. For this, we introduce a kinematic melody line extractor to encode speed information of the motion. Theoretically, we can either use the angular velocities or position velocities to encode the speed of a dance. In practice, we chose to use joint translation velocities for the purpose of motion control, similar to~\cite{kim2003rhythmic}. Inspired by \cite{fan2011example},  we extract the speed of the motion for just a few key joints, opposed to all of them. Specifically, we focused on the left and right shoulders, elbows, hands, knees, feet, and the head. From these joints, we can express the salient kinematic movement, along with
the importance of the positions in motion features~\cite{lee2002interactive}. The speed of motion is determined by the change in position between neighboring poses. Specifically, the speed of the motion at frame $t$ is the sum of speed of each key joint $k$. Mathematically speaking,
\begin{equation}
	L(t) = g(\sum_{k}\left | p_{t}^{k}-p_{t-1}^{k} \right |).
	  \label{equ:melody_line}
\end{equation}
where the $p_{t}^{k}$ is the position of joint $k$ at frame $t$.
To ensure smoothness, we use a Gaussian filter $g$ to smooth the motion speed to get the motion melody line $L(t)$. Thus, $L(t)$ (\ie the motion speed) is a strong signal (Fig.~\ref{fig:dance_melody_line}). Meanwhile, the signal is highly coupled due to summing of speeds-- different motions may have the same melody line. 

Since dance focuses on melody changes, we use the melody change trend as the control signal instead of directly using the melody line.
Specifically, we extract the melody line relative to the value of current frame in a one second interval, and use it as the melody control signal. We represent the melody control signal $c_{t}^{l}$ by sparsely sampling in the temporal domain. Hence, starting from frame $t$ of the melody line clip, a one second sequence of future frames is extracted (\ie 60 frames from our 60 FPS dance motion). Then, we down-sampled in the temporal domain to 12-1D points. Finally, we subtract the speed of frame $t$ to produce the change in speed and, thus, the melody control signal, 
\begin{equation}
	c_{t}^{l} = \left [ L(t+t_{int})- L(t), ..., L(t+n*t_{int})- L(t), ...  \right ],
	  \label{equ:melody_control}
\end{equation}
where $t_{int}$ is the sampling interval and $n$ is the sampling index.

\subsection{Motion representation}\label{sec:motion_representation}
Human motion is modeled as an articulated figure with rigid links (\ie socket joints) connected ball-to-ball. Each frame is represented by the root translation ($p_{x}, p_{y}, p_{z}$), rotation ($ r_{x}, r_{y}, r_{z}$), and the other joint rotations with respect to the parents (\ie $ r_{jx}, r_{jy}, r_{jz}$, and $j$ is the joint index). However, such a motion representation is a relative feature with local information related their parents. In addition, the translation and rotation of the root joint is relative to world coordinates. Ultimately, this increases the motion feature space, which increases the modelling complexity. Thus, we adopt the relative translation and rotation of root joint and add 3D joint positions, angle, and position velocities to the representation. This allows us to better model human motions. In our method, the motion feature at frame $t$ is represented as the rotations using quaternion exponential mapping $x_{t}^{r}$~\cite{grassia1998practical}, the angular velocities $x_{t}^{\omega }$, the 3D joint positions $x_{t}^{p}$, the joint linear velocities $x_{t}^{v}$, and foot contact information $x_{t}^{f}$. For the rotation of the root joint, we use the relative rotation $\Delta r_{y}$ of the current and previous frames (\ie the rotation about the Y-axis), and the $x$ and $z$ translations of the root joint are defined on the local coordinates of the previous frame ($\Delta p_{x}, \Delta p_{z}$), like in~\cite{holden2016deep}. In summary,

\begin{equation}
x_{t}=(x_{t}^{r}, x_{t}^{\omega }, x_{t}^{p}, x_{t}^{v}, x_{t}^{f})
\label{equ:motion_feature}
\end{equation}
\begin{equation}
x_{t}^{r}=({[ r_{t,x}, \Delta r_{t,y}, r_{t,z},}\\ 
{..., r_{t,jx}, r_{t,jy}, r_{t,jz}]})\\
\label{equ:motion_rot}
\end{equation}
\begin{equation}
x_{t}^{p}=(\Delta p_{t}, t_{t,y}, \Delta p_{t,z}, ..., p_{t,jx}, p_{t,jy}, p_{t,jz}).
\label{equ:motion_position}
\end{equation}
We extract motion features from two aspects (\ie rotation and position) to maximize the amount of motion information. Then, we add the angular velocity and linear velocity to more fully represent the motion feature. In addition, the information about the foot contact $x_{t}^{f}$ is added to reduce the foot sliding  in the generated frames. Like in~\cite{holden2017phase,zhuang2020music2dance}, the foot contact labels(ground-truth) are detected by the height and speed of $toe_{end}$ per frame.

%% file: model.tex
\section{Generative Model}\label{sec:model}
Next, we introduce the structure of the proposed model, and then we discuss the training details.

\subsection{Encoder-decoder structure}\label{sec:model_structure}
Our end-to-end framework is shown in Figure \ref{fig:teaser}: the proposed models the \gls{pdf} of the predicted motion conditioned on control signals $c_{t}=\left [ c_{t}^{l}, c^{s} \right ]$ as
\begin{equation}
	Pr({x}| c^{l}, c^{s})=\prod_{t=1}^{T}Pr({x}_{t}|X_{pre}, c_{t}^{l}, c^{s}),
	  \label{equ:model_pdf}
\end{equation}
\begin{equation}
Pr({x}_{t}|X_{pre}, c_{t}^{l}, c^{s})=\Gamma (X_{pre}, c_{t}^{l}, c^{s})=\Gamma _{D}(\Gamma _{E}(X_{pre}, c_{t}^{l}, c^{s})),
	  \label{equ:model_curframe}
\end{equation}
where $X_{pre}={x}_{t-k-1},\dots,{x}_{t-1}$ is the motion of the previous $k$ frames and $\Gamma$ is our model made-up of encoder $\Gamma _{E}$ and decoder $\Gamma _{D}$.

\noindent\textbf{Encoder.} Inspired by~\cite{zhuang2020music2dance}, we adopt temporal dilated convolution to extract input features for improved robustness to noise. The motion controllable feature $m_{t}$ is extracted during encoding via
\begin{equation}
m_{t} = \Gamma _{E}(X_{pre}, c_{t}^{l}, c^{s})=\sum_{i}\Gamma _{E}^{r,i}(\Gamma _{E}^{e}(X_{pre}), c_{t}^{l}, c^{s}).
\label{equ:model_encoder}
\end{equation}
Due to the high complexity of motion data, we first encode the input with a two stacked \emph{Conv1D+Relu} module $\Gamma _{E}^{e}$. Then, the residual control module $\Gamma _{E}^{r,i}$ fuses motion coded features and control signals similar to \cite{zhuang2020music2dance}. We stack 10-$\Gamma _{E}^{r,i}$ and sum the outputs to produce motion controllable feature $m_{t}$.
Furthermore, the dilated convolution in $\Gamma _{E}^{r,i}$ can extract the temporal fusion information of the motion sequence by increasing the receptive field.
Also, $\Gamma _{E}^{r,i}$ accepts the control signals as input to the 1D-conv layer with a kernel size of 1. Then, the fused temporal motion feature is fused with the control features by summing. However, the coupling formed via addition lacks in strength, which allows us to swap the control signals from training-to-inference. In our experiments, the input channel of $\Gamma _{E}^{r,i}$ is 128D, while the motion controllable feature $m_{t}$ is 512D.

\begin{figure}
	\centering
	\includegraphics[width=\linewidth]{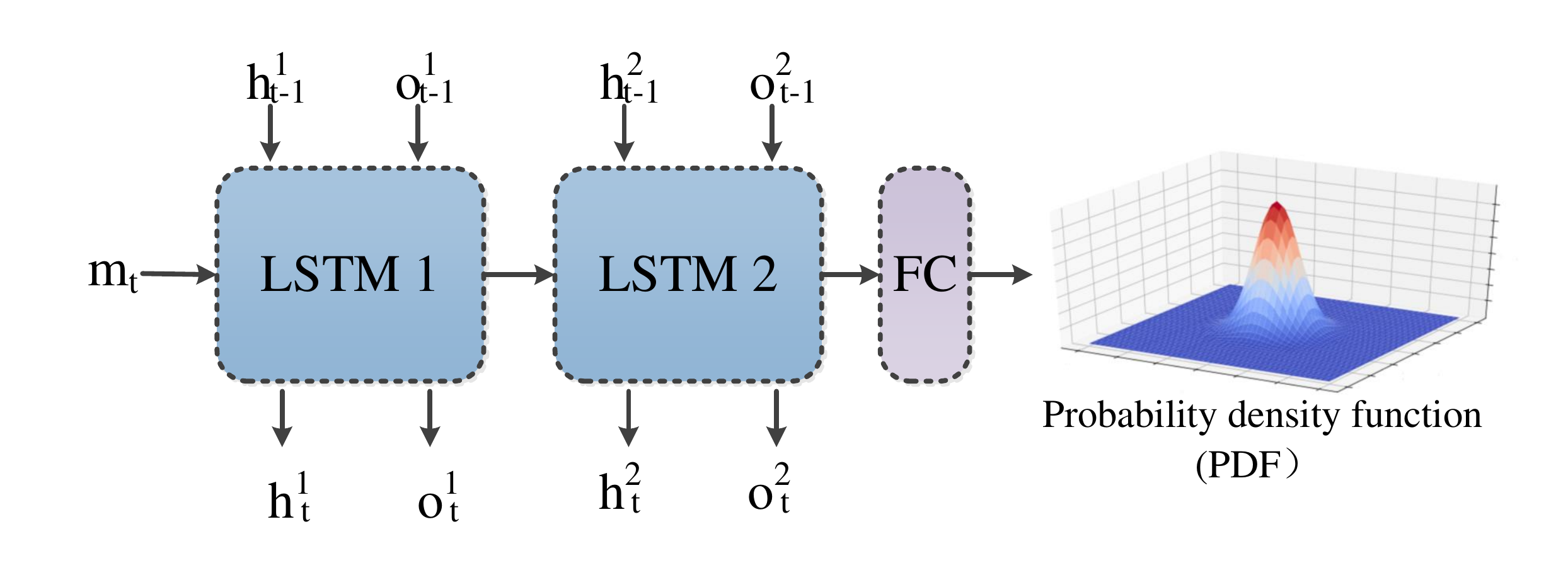}
	\caption{\textbf{The decoder of our model}.}
	\label{fig:model_decoder}
\end{figure}

\vspace{1mm}
\noindent\textbf{Decoder.} To improve the temporal correlation of the output motion, we use two \glspl{lstm} and a fully connected layer to decode $m_{t}$ and to predict the \gls{pdf} of current frame (Fig.~\ref{fig:model_decoder}). Mathematically speaking,
\begin{equation}
Pr({x}_{t}|{m}_{t}),{h}_{t}^{1},{o}_{t}^{1}, {h}_{t}^{2},{o}_{t}^{2} = \Gamma _{D}({m}_{t},{h}_{t-1}^{1},{o}_{t-1}^{1}, {h}_{t-1}^{2},{o}_{t-1}^{2})
\label{equ:model_decoder1}
\end{equation}
where ${h}_{t}^{1}$ and  ${o}_{t}^{1}$ represent the hidden state and cell memory of the first \gls{lstm} layer, respectively, and ${h}_{t}^{2},{o}_{t}^{2}$ are the second layer (Fig.~\ref{fig:model_decoder}). The hidden states and cell memories of the \glspl{lstm} ensure temporal correlation between output frames such that the output motion is smooth and realistic.

\subsection{Training}\label{sec:loss}
\noindent\textbf{Training loss.} The output of \gls{tclstm} is the \gls{pdf} of frame $t$, which we model as a \gls{gmm},
with a loss defined as the negative log likelihood. Specifically,
\begin{equation}
L_{G}=-logPr(x_{t} | \omega _{t}^{i},{\mu }_{t}^{i},\Sigma _{t}^{i}) 
=-log\sum_{i=1}^{N}\omega _{i}\mathbb{N}(x_{t} | \mu _{t}^{i},\Sigma _{t}^{i}),
\label{equ:gmm_loss}
\end{equation}
\begin{equation}
	\omega _{t}^{i}=\frac{e^{\hat{\omega}_{t}^{i}}}{\sum e^{\hat{\omega}_{t}^{i}}},{\mu }_{t}^{i}=\hat{\mu }_{t}^{i},\Sigma _{t}^{i}=e^{\hat{\sigma }_{t}^{i}},
	  \label{equ:gmm_dis_para}
\end{equation}
where $N=1$, $\hat{\omega}$, $\hat{\mu }$, $\hat{\sigma }$ is the output, $\mu _{t}^{i}, \Sigma _{t}^{i}$ are the mean vector and co-variance matrix, respectively. Note that $\mu _{t}^{i}$ is the ground-truth motion frame. To ensure temporal smoothness of the output motion, the smoothness loss is optimized for just the mean vector via  $L_{S}=\sum_{t}(\hat{\mu }_{t+1}+\hat{\mu }_{t-1}-2\hat{\mu }_{t})$.
Note that the binary foot contact in $x_{t}$ is omitted from $\mu$. Instead, we use the binary cross entropy (BCE) loss to compute the foot contact loss as $L_{F}=BCE(\hat{x}_{t}^{f},x_{t}^{f})$, with
$x_{t}^{f}$ and $\hat{x}_{t}^{f}$ as the ground-truth and predicted foot contact, respectively.

In the end, our training loss can be described as a sum of losses:
\begin{equation}
	\L=L_{G}+ \lambda * L_{S}+ \beta * L_{F},
	  \label{equ:total_loss}
\end{equation}
where the balance parameters $\lambda=0.1$ and $\beta=1$.

Then, at inference, the generated motion frame can be obtained by sampling from the predicted \gls{pdf}.

\noindent\textbf{Implementation details.} 
To achieve dance synthesis with and without the \emph{dance melody line} control signal, we adopt a mix training strategy. That is, we pass the probabilistic input (\ie melody control signal) during training with a probability of 0.5. 
To obtain robustness, we apply data augmentation: (1) mirror transformations for additional dance motion, (2) added Gaussian noise (\ie $\mu_{noise}=0$ and $\sigma _{noise}=0.05$) to the input and ground-truth to learn to handle temporal accumulation error, and (3)
apply dropout (\ie 0.4) at the input to resolve the problems of over-fitting.
We initialize our model we use Xavier normal~\cite{glorot2010understanding}, and optimize via RMSprop~\cite{tieleman2012lecture}. Training runs for 500 epochs, starting with a learning rate of $4\times 10^{-4}$, and then dropping by a factor of 10 at epoch 300. The batch-size is 128, setting each sample as a motion sequence of 600 continuous frames. Our system is implemented using PyTroch 1.2 on a PC with Intel I7 CPU, 32G RAM, and a GeForce-GTX 1080Ti.

%% file: experiment.tex
\section{Experiment}\label{sec:Experiment}

\begin{figure*}
	\centering
	\includegraphics[width=.95\linewidth]{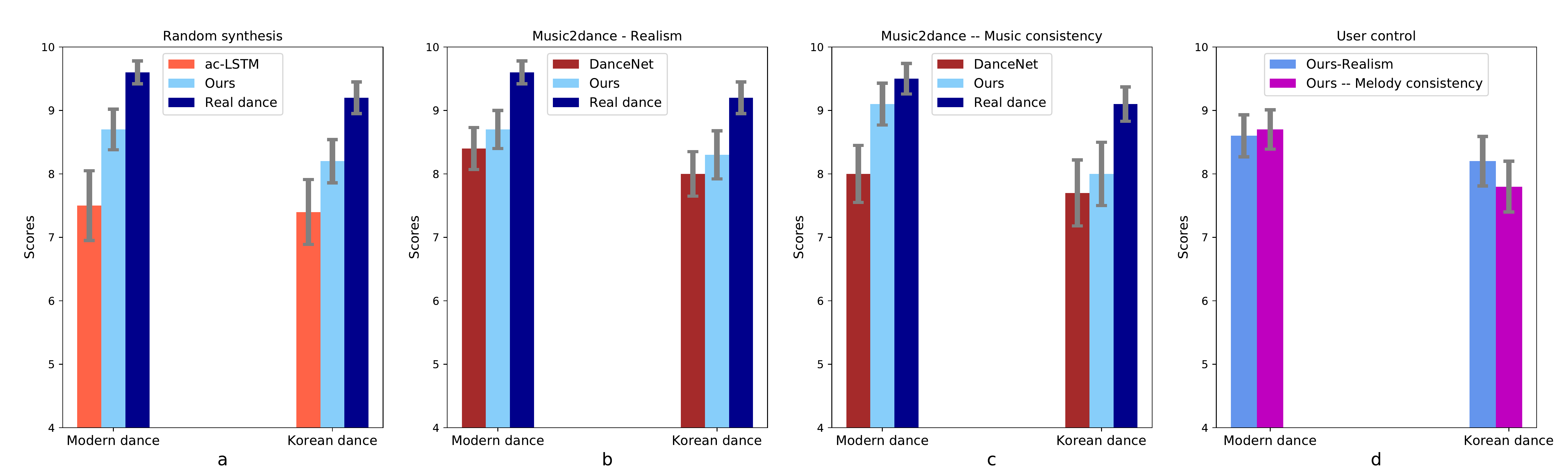}
	\caption{\small{The user studies of different applications: (a) Random synthesis, we asked 10 users to score the synthesized dances of acLSTM, our method(no melody control signal), and the real dances(the dances in dataset); (b) Music2Dance (realism), we asked 10 users to score the realism of synthesized dances of DanceNet, our method(with music melody line), and the real dances; (c) Music2Dance (consistency), score the music-consistency of synthesized dances; (d) User control, score the realism and melody-consistency of synthesized dances of our method(user given melody line).}}
	\label{fig:user_study}
\end{figure*}

The proposed melody control signal allows it to be toggled on and off for different applications (\ie random synthesis, Music2Dance, and user control). As far as we know, this is the first dance motion synthesizer proposed for a wide range of applications with the same model(the same parameters). Furthermore, this is the first user-controlled dance motion synthesizer for specific motions. 

In this section, we describe each application, and compare with \gls{sota} methods (Table~\ref{table:app_compare}). The \gls{sota} random synthesis model (\ie the ac-\gls{lstm} \cite{li2017auto}), adopts a unique strategy (\ie auto-condition) to train the \gls{lstm}. However, it lacks controllability for dance motion. Tang~\etal~proposed an \gls{ae}-based \gls{lstm} (\ie \gls{lstm}-\gls{ae}) to synthesize dance motion from music~\cite{tang2018dance}, but its synthesized dances are unrealistic and out of sync with the music. Furthermore, it lacks an ability to be used in other applications. DanceNet was proposed to synthesize dance from music~\cite{zhuang2020music2dance}. However, it lacks the ability to synthesize a variety of dance motions from the same music. Our model synthesizes realistic and diverse dance motions of different dance types, and  the dance motions synthesized from same music are various. The animations of all applications are shown in video demo as part of the supplemental material.

\setlength{\tabcolsep}{4pt}
\begin{table}[t]
\begin{center}
\caption{Comparison of realism (FID; lower is better) and diversity (Diversity-\uppercase\expandafter{\romannumeral1}: synthesized conditioned different initial frames, Diversity-\uppercase\expandafter{\romannumeral2}: synthesized conditioned same initial frames; higher is better).
}
\label{table:random_compare}
\scriptsize
\begin{tabular}{l c c c | c c c}
\toprule
   & \multicolumn{3}{c}{Modern Dance} &  \multicolumn{3}{c}{Korean Dance}\\ 
   \cmidrule(lr){2-4}\cmidrule(lr){5-7}
   & FID & Diversity-\uppercase\expandafter{\romannumeral1} & Diversity-\uppercase\expandafter{\romannumeral2} & FID & Diversity-\uppercase\expandafter{\romannumeral1} & Diversity-\uppercase\expandafter{\romannumeral2}\\
Real Dances & 6.5 & 55.4 & -- & 5.6 & 42.5 & -- \\
\midrule
ac-\gls{lstm}\cite{li2017auto} & 23.4 & 41.1  & 7.8 & 22.5 & 28.9  & 7.5\\
\midrule
Ours(w/o \gls{lstm} decoder) &  15.6 & 48.1  & 37.9 & 11.7 & 36.2 & 24.8 \\
Ours & \textbf{10.6} & \textbf{50.9}  & \textbf{40.1} & \textbf{7.4} & \textbf{38.5} & \textbf{26.3} \\
\bottomrule	
\end{tabular}
\end{center}
\end{table}
\setlength{\tabcolsep}{1.4pt}

\begin{figure}
	\centering
	\includegraphics[width=\linewidth]{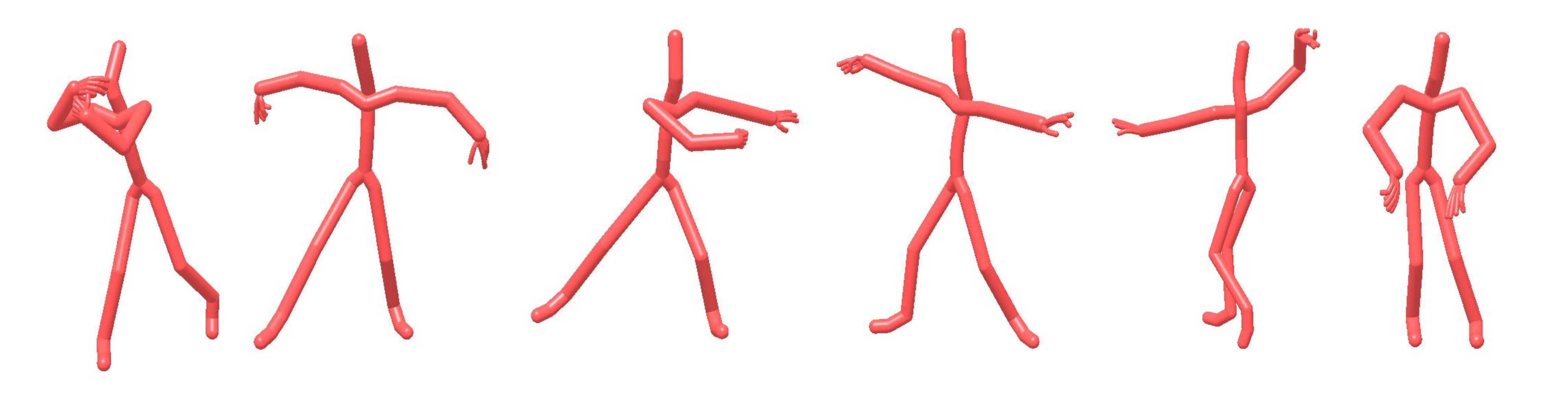}
	\caption{Example of random synthesis result.}
	\label{fig:result_random_synthesis}
\end{figure}

\subsection{Random synthesis}\label{sec:random_synthesis}
Given the dance type and initial dance motion frames(30 frames), our model can generate realistic and diverse dance sequences with its own style. With the same initial input frames, our model can generate different dance motion sequences, as shown in Figure~\ref{fig:result_random_synthesis}. Since the predicted motion frame needs to be sampled from the\gls{pdf}of the model output, which can effectively increase motion diversity. Then the sampled motion frame is fed back to the input to generate follow-up frame. To demonstrate our method, we compared with the \gls{sota} model: ac-\gls{lstm} \cite{li2017auto}. We randomly synthesize 15 dance motion sequences for each dance type, and every three synthesized sequences share the same initial input frames. We evaluate the random synthesized dance motion from realism and diversity. The realism can be evaluated by Fr\'echet Inception Distance(FID)\cite{heusel2017gans}, similar to \cite{yan2019convolutional,lee2019dancing}. We adopt 3 temporal convolution layers and 1 Bi-\gls{lstm} layer as a feature extractor to obtain the FID score since the FID needs an action classifier to extract dance features. Our method can synthesize diverse sequences with the different/same initial frames, so we can evaluate the diversity by the dance features extracted by the action classifier, that is, the average feature distance among different sequences. Diversity-\uppercase\expandafter{\romannumeral1} evaluates the diversity of synthesized different dance sequences conditioned on different initial frames, Diversity-\uppercase\expandafter{\romannumeral2} evaluates the diversity conditioned on same initial frames, and the result is shown in Table \ref{table:random_compare}. For better comparison, we use user study to score the realism and diversity of dance(10 users), as shown in Figure~\ref{fig:user_study}a. Our model can synthesize more realistic dance motion sequences, close to real dance sequences. It is worth noting that the dances generated by our model are diverse, while ac-\gls{lstm} \cite{li2017auto} can not synthesize diverse dances at all for the same initial frames. Their method can only synthesize the same dance with the same initial frames. One explanation is that their training strategy 
avoids the temporal accumulation error, but the model completely loses diversity(just overfit to the train data). Our method can achieve the random synthesis of diverse motion sequences. Because our model has complex and robust modelling capabilities and the output of our model is the probabilistic density (we need to sample for predicted frame), which increases dance diversity.

\subsection{Music2Dance}\label{sec:Music2Dance}
Synthesizing music-consistent dance is an interesting and challenging task. It requires that the synthesized dance can be consistent with the music rhythm, style and melody. However, music and dance are weakly related, and it does not determine the specific dance posture, that is, music does not determine whether the dance moves are leg lifting, jumping, or circling. Therefore, how to establish a correlation between music and dance is very important. Tang~\etal~\cite{tang2018dance} synthesized dance directly from music, and did not explicitly establish the relationship between music and dance, so the synthesized dance is not realistic. Zhuang~\etal~\cite{zhuang2020music2dance} added music feature in the process of auto-regressive synthesis motion, but did not establish the relationship between music and dance.

How to determine the relationship between music and dance is the core difficulty of music2dance task. From professional choreographers, we know the relationship between music and dance is reflected in the melody and rhythm. Therefore, we construct the relationship between music and dance through the melody line of music and dance. In section \ref{sec:dance_melody_line}, we propose the dance melody line to express the melody and rhythm of dance. In order to extract the music melody line, we introduce a simple and effective extraction method: extract onset strength by librosa~\cite{mcfee2015librosa} or madmom~\cite{bock2016madmom} and then smooth it through a Gaussian filter. 

\begin{figure}[t!]
	\centering
	\includegraphics[width=.95\linewidth]{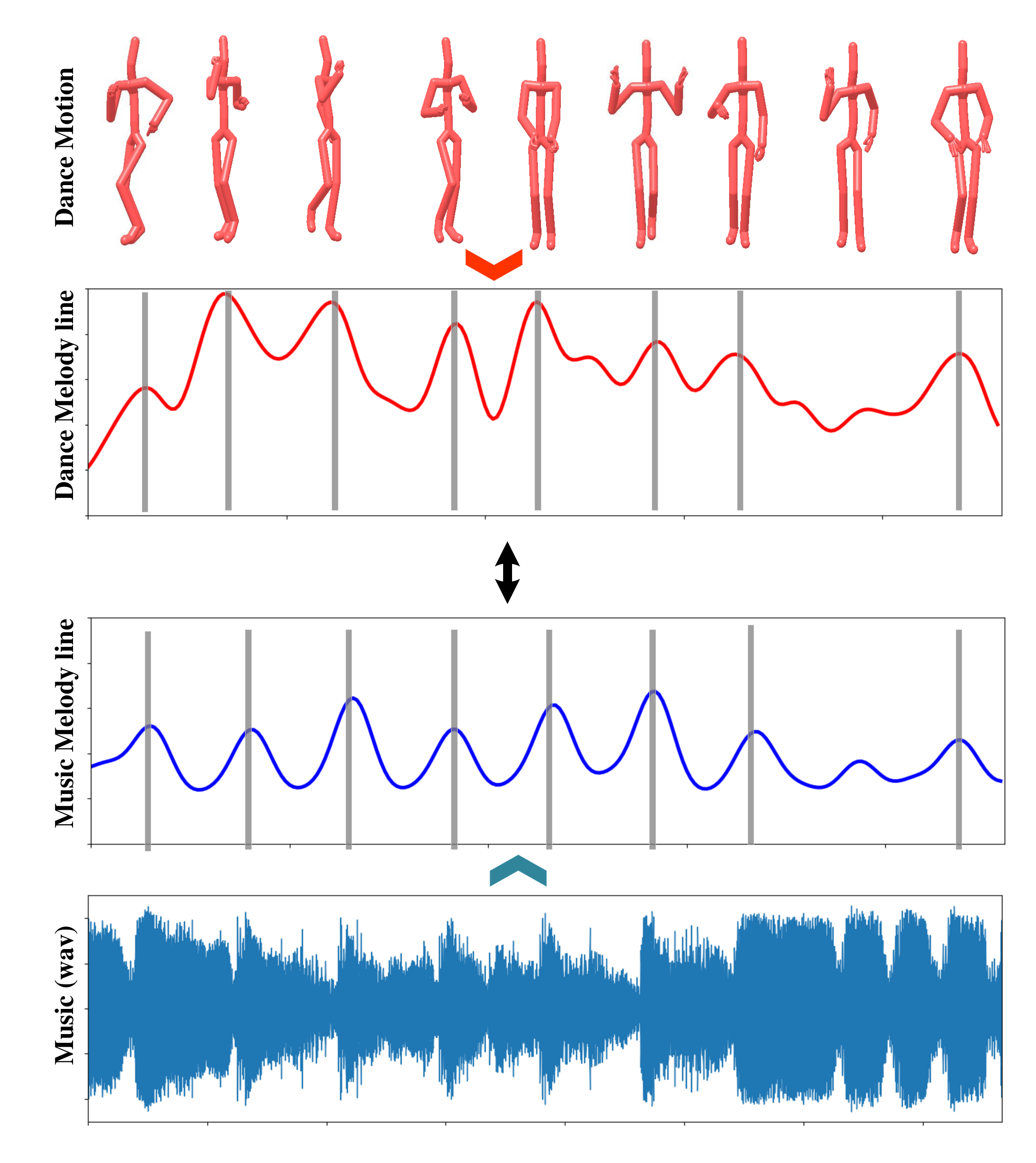}
	\caption{\textbf{The melody lines of dance and music in music-dance pair dataset}.}
	\label{fig:music_dance_melody_line}
\end{figure}

We select a segment from the music-dance pair data to obtain the melody lines of music and dance, and compare the relationship between them, as shown in Figure~\ref{fig:music_dance_melody_line}. Although the melody value of each frame is not necessarily the same, the change trend and peak value of the two melody lines are basically the same, indicating that the consistency of the melody rhythm between music and dance can be reflected through the melody line. In section \ref{sec:dance_melody_line}, we introduce that the melody control signal of the model is the change trend of the melody line relative to the current frame, so we can directly adopt the change trend of the music melody line as the melody control signal.

\setlength{\tabcolsep}{4pt}
\begin{table*}[t]
\begin{center}
\caption{Comparison of realism (FID), diversity (Diversity-\uppercase\expandafter{\romannumeral1}: synthesized conditioned different initial frames and different melody line, Diversity-\uppercase\expandafter{\romannumeral2}: synthesized conditioned same initial frames and different melody line, Diversity-\uppercase\expandafter{\romannumeral3} (Multi-modal): synthesized conditioned same initial frames and same melody line), rhythm-consistent (rhythm hit rate, higher is better).
}
\label{table:music2dance_comparison_result}
\small
\begin{tabular}{l c c c c c | c c c c c}

\toprule
   & \multicolumn{5}{c}{Morden Dance} &  \multicolumn{5}{c}{Korean Dance}\\ 
   \cmidrule(lr){2-6}\cmidrule(lr){7-11}
   & FID & Diversity-\uppercase\expandafter{\romannumeral1} & Diversity-\uppercase\expandafter{\romannumeral2} & Diversity-\uppercase\expandafter{\romannumeral3} &   Rhythm & FID & Diversity-\uppercase\expandafter{\romannumeral1} & Diversity-\uppercase\expandafter{\romannumeral2} & Diversity-\uppercase\expandafter{\romannumeral3} &   Rhythm \\
Real Dances & 6.5 & 55.4 & -- & -- & 57.9\%  & 5.6 & 42.5 & -- & --  & 68.3\%\\
\midrule
\gls{lstm}-\gls{ae}\cite{tang2018dance} & 81.3 & 12.4 & 9.4 & 8.9 & 13.6\%  & 75.6 & 10.2 & 7.9 & 7.6  & 15.1\%\\
DanceNet\cite{zhuang2020music2dance} & 15.2 & 49.3 & 40.1 & 7.6 & \textbf{56.7\%}  & 10.4 & 36.3 & 30.2 & 6.5  & 64.3\%\\
\midrule
Ours(w/o \gls{lstm}s decoder) & 13.8 & 50.3 & \textbf{50.1} & 42.5 & 55.4\%  & 10.1 & 35.3 & \textbf{32.7} & 24.9  & \textbf{67.3}\%\\
Ours & \textbf{11.2} & \textbf{50.8} & 48.9 & \textbf{43.4} & 56.2\%  & \textbf{7.8} & \textbf{36.5} & 31.9 & \textbf{26.4}  & 66.8\%\\

\bottomrule	
\end{tabular}
\end{center}
\end{table*}
\setlength{\tabcolsep}{1.4pt}

\begin{figure}
	\centering
	\includegraphics[width=.95\linewidth]{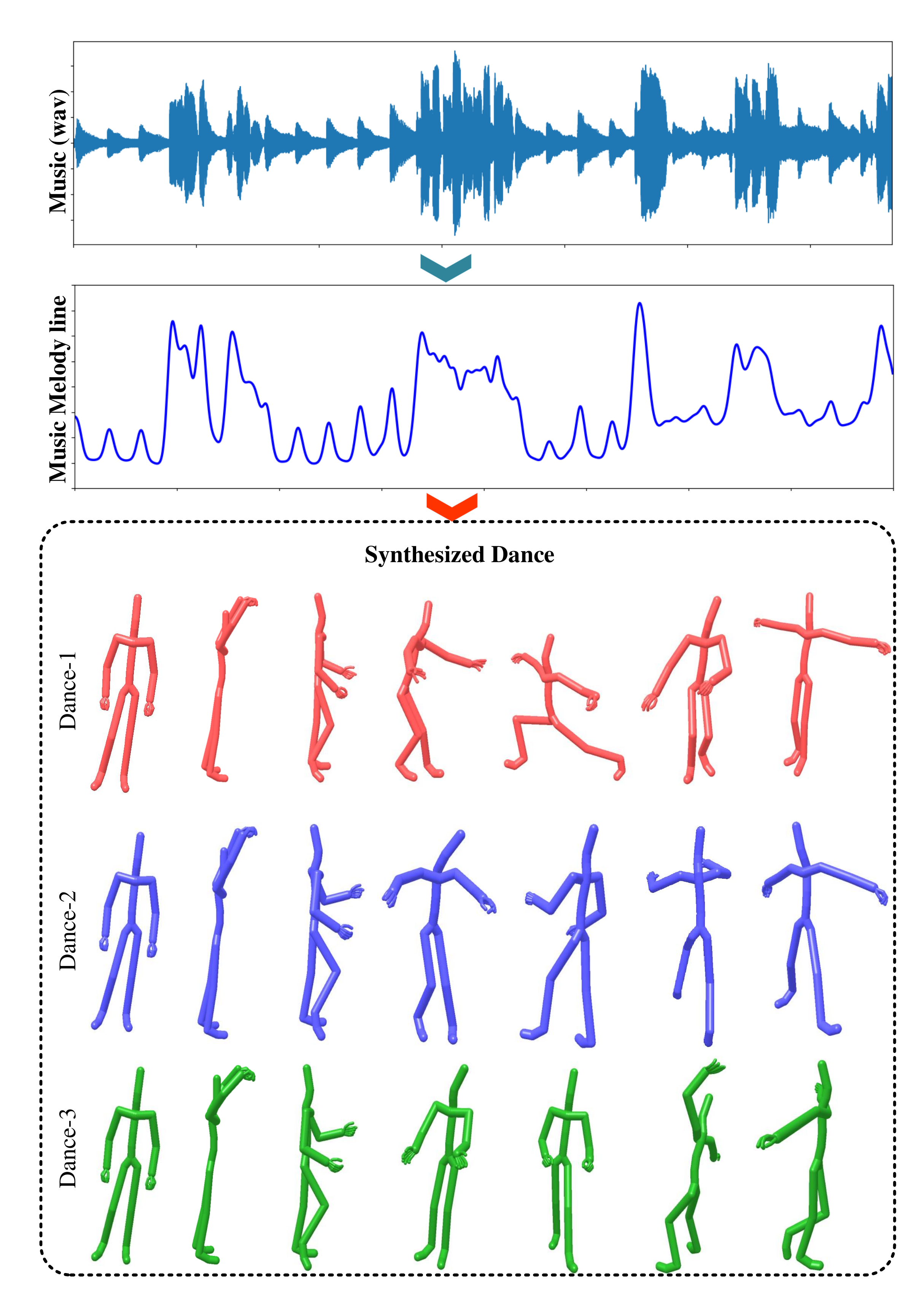}
	\caption{\textbf{Music2Dance result}. Three dance sequences are synthesized conditioned on same initial frames and same music(melody line).}
	\label{fig:result_music2dance_synthesis}
\end{figure}

To demonstrate the effectiveness of our approach, we compare with two \gls{sota} methods, \gls{lstm}-\gls{ae}~\cite{tang2018dance} and DanceNet \cite{zhuang2020music2dance}. We use the FID to evaluate the dance realism, and the average feature distance to evaluate the dance diversity. For this application, dance diversity is evaluated from three views: (1) Diversity-\uppercase\expandafter{\romannumeral1}, the diversity of the different synthesized dance sequences conditioned on different initial frames and different melody line; (2) Diversity-\uppercase\expandafter{\romannumeral2}, the diversity is conditioned on same initial frame and different melody lines; and (3) Diversity-\uppercase\expandafter{\romannumeral3}, the diversity is conditioned on same initial frame and the same melody line. Diversity-\uppercase\expandafter{\romannumeral3} reflects the multi-modality power of Music2Dance. In addition, we use the rhythm consistency (\ie the rhythm hit rate) to evaluate our methods like in~\cite{lee2019dancing,zhuang2020music2dance}. 
We randomly select 5 initial sequences to synthesize 30 dance motion sequences for each dance type using the two methods mentioned and our approach. For each, three were synthesized conditioned on the same initial sequence and the same melody line, and three were synthesized conditioned on the same initial sequence and a different melody line. The quantitative results are shown in Table~\ref{table:music2dance_comparison_result}. We conduct a user study to evaluate the realism of the music-consistency (Fig.~\ref{fig:user_study}, b \& c).  Our results significantly outperform \gls{lstm}-\gls{ae}. We believe that directly mapping music to get the dance movement is unreasonable due to the weak correlation between dance and music. DanceNet \cite{zhuang2020music2dance} uses music features as the conditions to synthesize dance, but this method directly inputs music features without explicitly analyzing the correlation between music and dance. So, the model takes a long time to train (\ie >1,000 epochs). However, the consistency in the synthesized dance and the music is low. In addition, the output of DanceNet is a probability distribution - a model trained for too long would cause it to collapse (\ie it lacks in diversity), and especially the synthesized dance sequences conditioned on same initial frame. We explicitly analyze the relationship between music and dance represented by the melody lines. During training, the dance melody line is used as the control signal. Then, at inference, the music melody line is used as the control signal. This strategy ensures the controllable effect (\ie music consistency) of the synthesized dance. More importantly, the highly coupled of the melody lines and low training difficulty (\ie about 500 epochs), our method prevents the model from collapsing to synthesize various dances.

In addition to music in WAV format, our method also works with audio in other formats (\eg MIDI, an electronic music format): melody lines are captured as a temporal sequence, as it is easier to obtain melody line (\ie obtained from the change of music note). 

\begin{figure}
	\centering
	\includegraphics[width=.95\linewidth]{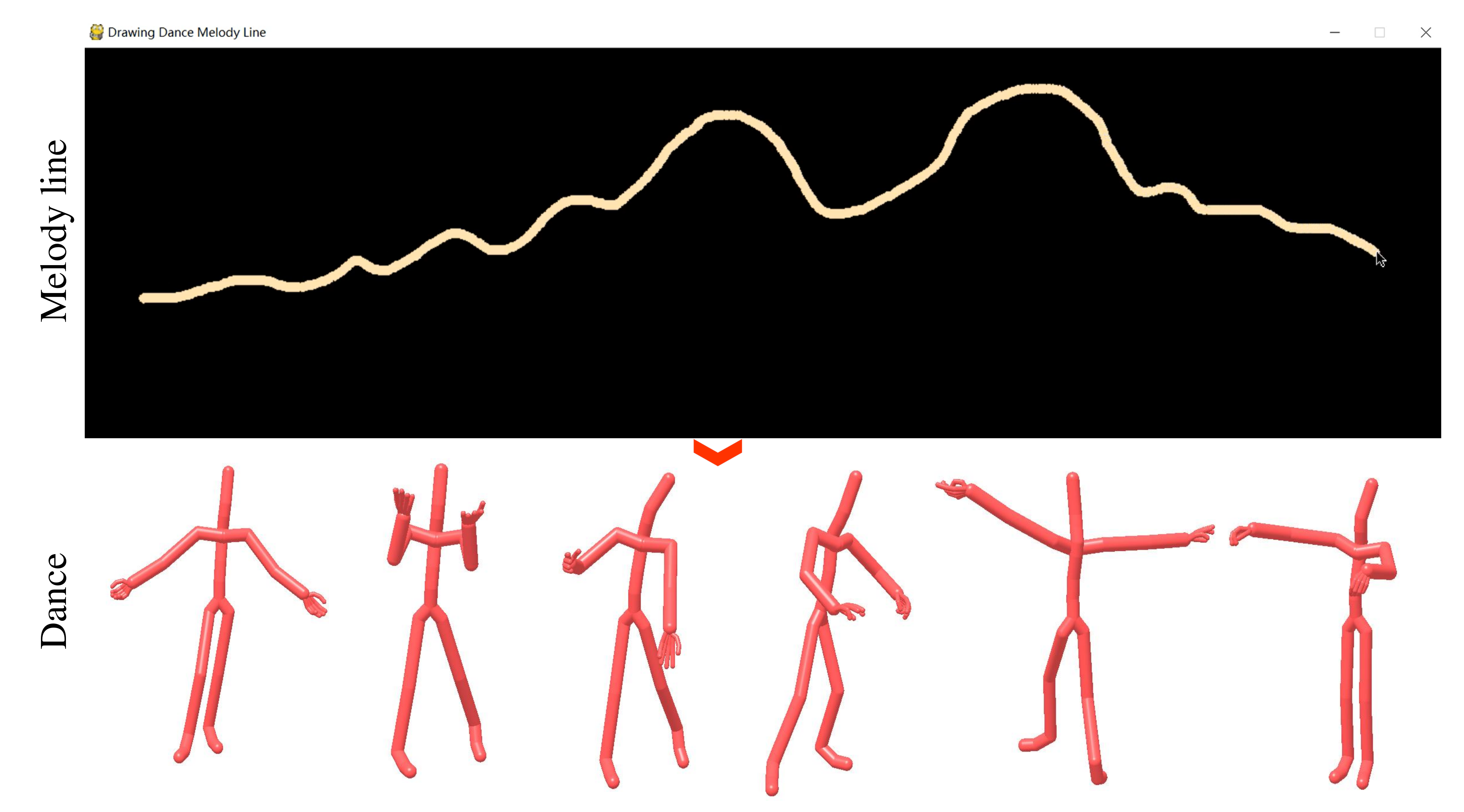}
	\caption{User controls - the melody line is drawn by users.}
	\label{fig:user_control_result}
\end{figure}

\subsection{User control}\label{sec:user_control}
Given the dance type, our model synthesizes realistic, diverse dance motions conditioned on the melody line. The melody line is a simple 1D signal (Fig.~\ref{fig:dance_melody_line},~\ref{fig:music_dance_melody_line}, and~\ref{fig:result_music2dance_synthesis}). Thus, allowing an ordinary user to create dances, opposed to depending on a professional choreographer - there is a variety of ways  the melody line can be described: \eg drawing (Fig.~\ref{fig:user_control_result}). We built an end-to-end system based on two steps: (1) the user draws the melody line via mouse inputs and (2) the synthesizer generates a dance according to the melody line. Note that since the lines drawn by the user are evenly sampled as melody lines, our model synthesizes melody-consistent dance motions. 
    
In the end, this is the first model with such controls for synthesizing dances. Thus, we evaluate our method by user study. We asked 10 users draw 3 melody lines. Then, they scored each dance sequence separately by measuring dance realism and melody-consistency (Fig.~\ref{fig:user_study} d). Our method synthesizes realistic dances, while ensuring melody-consistency, especially for the modern dance.


\subsection{Discussion and future work}\label{sec:dis_future}

\noindent\textbf{Ablation study.} We propose the \gls{tclstm} model, which is composed of two parts: an encoder (temporal convolution) and a decoder (\gls{lstm}). In order to verify the ability of our model, we conducted a comparative experiment. We directly use the temporal convolution model (without \gls{lstm}) to model the dance, and adopt same training strategy. We perform quantitative evaluations on two different applications, random synthesis and Music2Dance (Table~\ref{table:random_compare} and~\ref{table:music2dance_comparison_result}). When the \gls{lstm}s is omitted from the decoder, the synthesized dances worsen, and especially the in realism (\ie FID), which shows that building temporal dependencies via \gls{lstm}s in decoder improves modeling capabilities.

\vspace{1mm}
\noindent\textbf{Result discussion.} Our method realizes different dance synthesis applications (\ie more than the three applications described above). For example, musical notation synthesis dance (\ie to extract a melody line from musical notation). For these applications, we found a worthwhile phenomenon in the experiment. That is, the melody-consistency (\ie controllable effect) of modern dance is superior to that in Korean dance; also, the diversity of Korean dance is inferior to modern dance. The first reason lies in the dance data. The dance steps of Korean dance in dataset are inconsistently distributed, which makes it difficult to model such unevenly distributed data. The second reason is that the rhythm and melody of Korean dance is very fast, which causes poor temporal smoothness and high dynamic complexity.

\vspace{1mm}
\noindent\textbf{Future work.} 
To the best of our knowledge, we are the first to propose a generative model with controllable dance synthesis via the simple and effective use of 1D control signals. However, there are some topics worth discussing: \ie how to quantitatively evaluate the controls effectiveness, mediocre modeling ability for fast-rhythm Korean dance (mentioned above), the foot sliding (it is difficult to solve by IK for complex dance motion). The topics raised here are subject of future work.

%% file: conclusion.tex
\section{Conclusion}\label{sec:conclusion}
We introduced a novel generative model, \glspl{tclstm}. Based on temporal \glspl{cnn} and \glspl{lstm}, \gls{tclstm} synthesizes realistic, diverse dances (\ie motion sequence). Our model can handle different dance types for various applications: random synthesis, music2dance, and user control. We demonstrated quantitative results and user studies establishing the effectiveness of our method, and our model can synthesize more realistic and diverse dance motion sequences, achieving state-of-the-art results.

%% file: main.bbl

\begin{thebibliography}{40}


\ifx \showCODEN    \undefined \def \showCODEN     #1{\unskip}     \fi
\ifx \showDOI      \undefined \def \showDOI       #1{#1}\fi
\ifx \showISBNx    \undefined \def \showISBNx     #1{\unskip}     \fi
\ifx \showISBNxiii \undefined \def \showISBNxiii  #1{\unskip}     \fi
\ifx \showISSN     \undefined \def \showISSN      #1{\unskip}     \fi
\ifx \showLCCN     \undefined \def \showLCCN      #1{\unskip}     \fi
\ifx \shownote     \undefined \def \shownote      #1{#1}          \fi
\ifx \showarticletitle \undefined \def \showarticletitle #1{#1}   \fi
\ifx \showURL      \undefined \def \showURL       {\relax}        \fi
\providecommand\bibfield[2]{#2}
\providecommand\bibinfo[2]{#2}
\providecommand\natexlab[1]{#1}
\providecommand\showeprint[2][]{arXiv:#2}

\bibitem[\protect\citeauthoryear{B{\"o}ck, Korzeniowski, Schl{\"u}ter, Krebs,
  and Widmer}{B{\"o}ck et~al\mbox{.}}{2016}]%
        {bock2016madmom}
\bibfield{author}{\bibinfo{person}{Sebastian B{\"o}ck}, \bibinfo{person}{Filip
  Korzeniowski}, \bibinfo{person}{Jan Schl{\"u}ter}, \bibinfo{person}{Florian
  Krebs}, {and} \bibinfo{person}{Gerhard Widmer}.}
  \bibinfo{year}{2016}\natexlab{}.
\newblock \showarticletitle{Madmom: A new python audio and music signal
  processing library}. In \bibinfo{booktitle}{\emph{Proceedings of the 24th ACM
  international conference on Multimedia}}. ACM, \bibinfo{pages}{1174--1178}.
\newblock


\bibitem[\protect\citeauthoryear{Bowden}{Bowden}{2000}]%
        {bowden2000learning}
\bibfield{author}{\bibinfo{person}{Richard Bowden}.}
  \bibinfo{year}{2000}\natexlab{}.
\newblock \showarticletitle{Learning statistical models of human motion}. In
  \bibinfo{booktitle}{\emph{IEEE Workshop on Human Modeling, Analysis and
  Synthesis, CVPR}}, Vol.~\bibinfo{volume}{2000}.
\newblock


\bibitem[\protect\citeauthoryear{Boyd}{Boyd}{2004}]%
        {boyd2004dance}
\bibfield{author}{\bibinfo{person}{Jade Boyd}.}
  \bibinfo{year}{2004}\natexlab{}.
\newblock \showarticletitle{Dance, culture, and popular film}.
\newblock \bibinfo{journal}{\emph{Feminist Media Studies}} \bibinfo{volume}{4},
  \bibinfo{number}{1} (\bibinfo{year}{2004}), \bibinfo{pages}{67--83}.
\newblock


\bibitem[\protect\citeauthoryear{Brand and Hertzmann}{Brand and
  Hertzmann}{2000}]%
        {brand2000style}
\bibfield{author}{\bibinfo{person}{Matthew Brand} {and} \bibinfo{person}{Aaron
  Hertzmann}.} \bibinfo{year}{2000}\natexlab{}.
\newblock \showarticletitle{Style machines}. In
  \bibinfo{booktitle}{\emph{Proceedings of the 27th annual conference on
  Computer graphics and interactive techniques}}. ACM Press/Addison-Wesley
  Publishing Co., \bibinfo{pages}{183--192}.
\newblock


\bibitem[\protect\citeauthoryear{Chai and Hodgins}{Chai and Hodgins}{2005}]%
        {chai2005performance}
\bibfield{author}{\bibinfo{person}{Jinxiang Chai} {and}
  \bibinfo{person}{Jessica~K Hodgins}.} \bibinfo{year}{2005}\natexlab{}.
\newblock \showarticletitle{Performance animation from low-dimensional control
  signals}.
\newblock \bibinfo{journal}{\emph{ACM Transactions on Graphics (ToG)}}
  \bibinfo{volume}{24}, \bibinfo{number}{3} (\bibinfo{year}{2005}),
  \bibinfo{pages}{686--696}.
\newblock


\bibitem[\protect\citeauthoryear{Chai and Hodgins}{Chai and Hodgins}{2007}]%
        {chai2007constraint}
\bibfield{author}{\bibinfo{person}{Jinxiang Chai} {and}
  \bibinfo{person}{Jessica~K Hodgins}.} \bibinfo{year}{2007}\natexlab{}.
\newblock \showarticletitle{Constraint-based motion optimization using a
  statistical dynamic model}. In \bibinfo{booktitle}{\emph{ACM Transactions on
  Graphics (TOG)}}, Vol.~\bibinfo{volume}{26}. ACM, \bibinfo{pages}{8}.
\newblock


\bibitem[\protect\citeauthoryear{Fan, Xu, and Geng}{Fan et~al\mbox{.}}{2011}]%
        {fan2011example}
\bibfield{author}{\bibinfo{person}{Rukun Fan}, \bibinfo{person}{Songhua Xu},
  {and} \bibinfo{person}{Weidong Geng}.} \bibinfo{year}{2011}\natexlab{}.
\newblock \showarticletitle{Example-based automatic music-driven conventional
  dance motion synthesis}.
\newblock \bibinfo{journal}{\emph{IEEE transactions on visualization and
  computer graphics}} \bibinfo{volume}{18}, \bibinfo{number}{3}
  (\bibinfo{year}{2011}), \bibinfo{pages}{501--515}.
\newblock


\bibitem[\protect\citeauthoryear{Fragkiadaki, Levine, Felsen, and
  Malik}{Fragkiadaki et~al\mbox{.}}{2015}]%
        {fragkiadaki2015recurrent}
\bibfield{author}{\bibinfo{person}{Katerina Fragkiadaki},
  \bibinfo{person}{Sergey Levine}, \bibinfo{person}{Panna Felsen}, {and}
  \bibinfo{person}{Jitendra Malik}.} \bibinfo{year}{2015}\natexlab{}.
\newblock \showarticletitle{Recurrent network models for human dynamics}. In
  \bibinfo{booktitle}{\emph{Proceedings of the IEEE International Conference on
  Computer Vision}}. \bibinfo{pages}{4346--4354}.
\newblock


\bibitem[\protect\citeauthoryear{Glorot and Bengio}{Glorot and Bengio}{2010}]%
        {glorot2010understanding}
\bibfield{author}{\bibinfo{person}{Xavier Glorot} {and} \bibinfo{person}{Yoshua
  Bengio}.} \bibinfo{year}{2010}\natexlab{}.
\newblock \showarticletitle{Understanding the difficulty of training deep
  feedforward neural networks}. In \bibinfo{booktitle}{\emph{Proceedings of the
  thirteenth international conference on artificial intelligence and
  statistics}}. \bibinfo{pages}{249--256}.
\newblock


\bibitem[\protect\citeauthoryear{Grassia}{Grassia}{1998}]%
        {grassia1998practical}
\bibfield{author}{\bibinfo{person}{F~Sebastian Grassia}.}
  \bibinfo{year}{1998}\natexlab{}.
\newblock \showarticletitle{Practical parameterization of rotations using the
  exponential map}.
\newblock \bibinfo{journal}{\emph{Journal of graphics tools}}
  \bibinfo{volume}{3}, \bibinfo{number}{3} (\bibinfo{year}{1998}),
  \bibinfo{pages}{29--48}.
\newblock


\bibitem[\protect\citeauthoryear{Grochow, Martin, Hertzmann, and
  Popovi{\'c}}{Grochow et~al\mbox{.}}{2004}]%
        {grochow2004style}
\bibfield{author}{\bibinfo{person}{Keith Grochow}, \bibinfo{person}{Steven~L
  Martin}, \bibinfo{person}{Aaron Hertzmann}, {and} \bibinfo{person}{Zoran
  Popovi{\'c}}.} \bibinfo{year}{2004}\natexlab{}.
\newblock \showarticletitle{Style-based inverse kinematics}.
\newblock In \bibinfo{booktitle}{\emph{ACM SIGGRAPH 2004 Papers}}.
  \bibinfo{pages}{522--531}.
\newblock


\bibitem[\protect\citeauthoryear{Heusel, Ramsauer, Unterthiner, Nessler, and
  Hochreiter}{Heusel et~al\mbox{.}}{2017}]%
        {heusel2017gans}
\bibfield{author}{\bibinfo{person}{Martin Heusel}, \bibinfo{person}{Hubert
  Ramsauer}, \bibinfo{person}{Thomas Unterthiner}, \bibinfo{person}{Bernhard
  Nessler}, {and} \bibinfo{person}{Sepp Hochreiter}.}
  \bibinfo{year}{2017}\natexlab{}.
\newblock \showarticletitle{Gans trained by a two time-scale update rule
  converge to a local nash equilibrium}. In \bibinfo{booktitle}{\emph{Advances
  in neural information processing systems}}. \bibinfo{pages}{6626--6637}.
\newblock


\bibitem[\protect\citeauthoryear{Hewitt}{Hewitt}{2005}]%
        {hewitt2005social}
\bibfield{author}{\bibinfo{person}{Andrew Hewitt}.}
  \bibinfo{year}{2005}\natexlab{}.
\newblock \bibinfo{booktitle}{\emph{Social choreography: Ideology as
  performance in dance and everyday movement}}.
\newblock \bibinfo{publisher}{Duke University Press}.
\newblock


\bibitem[\protect\citeauthoryear{Holden, Komura, and Saito}{Holden
  et~al\mbox{.}}{2017}]%
        {holden2017phase}
\bibfield{author}{\bibinfo{person}{Daniel Holden}, \bibinfo{person}{Taku
  Komura}, {and} \bibinfo{person}{Jun Saito}.} \bibinfo{year}{2017}\natexlab{}.
\newblock \showarticletitle{Phase-functioned neural networks for character
  control}.
\newblock \bibinfo{journal}{\emph{ACM Transactions on Graphics (TOG)}}
  \bibinfo{volume}{36}, \bibinfo{number}{4} (\bibinfo{year}{2017}),
  \bibinfo{pages}{1--13}.
\newblock


\bibitem[\protect\citeauthoryear{Holden, Saito, and Komura}{Holden
  et~al\mbox{.}}{2016}]%
        {holden2016deep}
\bibfield{author}{\bibinfo{person}{Daniel Holden}, \bibinfo{person}{Jun Saito},
  {and} \bibinfo{person}{Taku Komura}.} \bibinfo{year}{2016}\natexlab{}.
\newblock \showarticletitle{A deep learning framework for character motion
  synthesis and editing}.
\newblock \bibinfo{journal}{\emph{ACM Transactions on Graphics (TOG)}}
  \bibinfo{volume}{35}, \bibinfo{number}{4} (\bibinfo{year}{2016}),
  \bibinfo{pages}{138}.
\newblock


\bibitem[\protect\citeauthoryear{Jain, Zamir, Savarese, and Saxena}{Jain
  et~al\mbox{.}}{2016}]%
        {jain2016structural}
\bibfield{author}{\bibinfo{person}{Ashesh Jain}, \bibinfo{person}{Amir~R
  Zamir}, \bibinfo{person}{Silvio Savarese}, {and} \bibinfo{person}{Ashutosh
  Saxena}.} \bibinfo{year}{2016}\natexlab{}.
\newblock \showarticletitle{Structural-RNN: Deep learning on spatio-temporal
  graphs}. In \bibinfo{booktitle}{\emph{Proceedings of the IEEE Conference on
  Computer Vision and Pattern Recognition}}. \bibinfo{pages}{5308--5317}.
\newblock


\bibitem[\protect\citeauthoryear{Kim, Park, and Shin}{Kim
  et~al\mbox{.}}{2003}]%
        {kim2003rhythmic}
\bibfield{author}{\bibinfo{person}{Tae-hoon Kim}, \bibinfo{person}{Sang~Il
  Park}, {and} \bibinfo{person}{Sung~Yong Shin}.}
  \bibinfo{year}{2003}\natexlab{}.
\newblock \showarticletitle{Rhythmic-motion synthesis based on motion-beat
  analysis}.
\newblock \bibinfo{journal}{\emph{ACM Transactions on Graphics (TOG)}}
  \bibinfo{volume}{22}, \bibinfo{number}{3} (\bibinfo{year}{2003}),
  \bibinfo{pages}{392--401}.
\newblock


\bibitem[\protect\citeauthoryear{Kovar, Gleicher, and Pighin}{Kovar
  et~al\mbox{.}}{2008}]%
        {kovar2008motion}
\bibfield{author}{\bibinfo{person}{Lucas Kovar}, \bibinfo{person}{Michael
  Gleicher}, {and} \bibinfo{person}{Fr{\'e}d{\'e}ric Pighin}.}
  \bibinfo{year}{2008}\natexlab{}.
\newblock \showarticletitle{Motion graphs}. In \bibinfo{booktitle}{\emph{ACM
  SIGGRAPH 2008 classes}}. ACM, \bibinfo{pages}{51}.
\newblock


\bibitem[\protect\citeauthoryear{Lau, Bar-Joseph, and Kuffner}{Lau
  et~al\mbox{.}}{2009}]%
        {lau2009modeling}
\bibfield{author}{\bibinfo{person}{Manfred Lau}, \bibinfo{person}{Ziv
  Bar-Joseph}, {and} \bibinfo{person}{James Kuffner}.}
  \bibinfo{year}{2009}\natexlab{}.
\newblock \showarticletitle{Modeling spatial and temporal variation in motion
  data}. In \bibinfo{booktitle}{\emph{ACM Transactions on Graphics (TOG)}},
  Vol.~\bibinfo{volume}{28}. ACM, \bibinfo{pages}{171}.
\newblock


\bibitem[\protect\citeauthoryear{Lee, Yang, Liu, Wang, Lu, Yang, and Kautz}{Lee
  et~al\mbox{.}}{2019}]%
        {lee2019dancing}
\bibfield{author}{\bibinfo{person}{Hsin-Ying Lee}, \bibinfo{person}{Xiaodong
  Yang}, \bibinfo{person}{Ming-Yu Liu}, \bibinfo{person}{Ting-Chun Wang},
  \bibinfo{person}{Yu-Ding Lu}, \bibinfo{person}{Ming-Hsuan Yang}, {and}
  \bibinfo{person}{Jan Kautz}.} \bibinfo{year}{2019}\natexlab{}.
\newblock \showarticletitle{Dancing to Music}. In
  \bibinfo{booktitle}{\emph{Advances in Neural Information Processing
  Systems}}. \bibinfo{pages}{3581--3591}.
\newblock


\bibitem[\protect\citeauthoryear{Lee, Chai, Reitsma, Hodgins, and Pollard}{Lee
  et~al\mbox{.}}{2002}]%
        {lee2002interactive}
\bibfield{author}{\bibinfo{person}{Jehee Lee}, \bibinfo{person}{Jinxiang Chai},
  \bibinfo{person}{Paul~SA Reitsma}, \bibinfo{person}{Jessica~K Hodgins}, {and}
  \bibinfo{person}{Nancy~S Pollard}.} \bibinfo{year}{2002}\natexlab{}.
\newblock \showarticletitle{Interactive control of avatars animated with human
  motion data}. In \bibinfo{booktitle}{\emph{Proceedings of the 29th annual
  conference on Computer graphics and interactive techniques}}.
  \bibinfo{pages}{491--500}.
\newblock


\bibitem[\protect\citeauthoryear{Lee, Lee, and Lee}{Lee et~al\mbox{.}}{2018}]%
        {lee2018interactive}
\bibfield{author}{\bibinfo{person}{Kyungho Lee}, \bibinfo{person}{Seyoung Lee},
  {and} \bibinfo{person}{Jehee Lee}.} \bibinfo{year}{2018}\natexlab{}.
\newblock \showarticletitle{Interactive character animation by learning
  multi-objective control}.
\newblock \bibinfo{journal}{\emph{ACM Transactions on Graphics (TOG)}}
  \bibinfo{volume}{37}, \bibinfo{number}{6} (\bibinfo{year}{2018}),
  \bibinfo{pages}{1--10}.
\newblock


\bibitem[\protect\citeauthoryear{Lee, Lee, and Park}{Lee et~al\mbox{.}}{2013}]%
        {lee2013music}
\bibfield{author}{\bibinfo{person}{Minho Lee}, \bibinfo{person}{Kyogu Lee},
  {and} \bibinfo{person}{Jaeheung Park}.} \bibinfo{year}{2013}\natexlab{}.
\newblock \showarticletitle{Music similarity-based approach to generating dance
  motion sequence}.
\newblock \bibinfo{journal}{\emph{Multimedia tools and applications}}
  \bibinfo{volume}{62}, \bibinfo{number}{3} (\bibinfo{year}{2013}),
  \bibinfo{pages}{895--912}.
\newblock


\bibitem[\protect\citeauthoryear{Li, Wang, and Shum}{Li et~al\mbox{.}}{2002}]%
        {li2002motion}
\bibfield{author}{\bibinfo{person}{Yan Li}, \bibinfo{person}{Tianshu Wang},
  {and} \bibinfo{person}{Heung-Yeung Shum}.} \bibinfo{year}{2002}\natexlab{}.
\newblock \showarticletitle{Motion texture: a two-level statistical model for
  character motion synthesis}. In \bibinfo{booktitle}{\emph{ACM transactions on
  graphics (ToG)}}, Vol.~\bibinfo{volume}{21}. ACM, \bibinfo{pages}{465--472}.
\newblock


\bibitem[\protect\citeauthoryear{Li, Zhou, Xiao, He, and Li}{Li
  et~al\mbox{.}}{2017}]%
        {li2017auto}
\bibfield{author}{\bibinfo{person}{Zimo Li}, \bibinfo{person}{Yi Zhou},
  \bibinfo{person}{Shuangjiu Xiao}, \bibinfo{person}{Chong He}, {and}
  \bibinfo{person}{Hao Li}.} \bibinfo{year}{2017}\natexlab{}.
\newblock \showarticletitle{Auto-conditioned lstm network for extended complex
  human motion synthesis}.
\newblock \bibinfo{journal}{\emph{arXiv preprint arXiv:1707.05363}}
  \bibinfo{volume}{3} (\bibinfo{year}{2017}).
\newblock


\bibitem[\protect\citeauthoryear{Martinez, Black, and Romero}{Martinez
  et~al\mbox{.}}{2017}]%
        {martinez2017human}
\bibfield{author}{\bibinfo{person}{Julieta Martinez},
  \bibinfo{person}{Michael~J Black}, {and} \bibinfo{person}{Javier Romero}.}
  \bibinfo{year}{2017}\natexlab{}.
\newblock \showarticletitle{On human motion prediction using recurrent neural
  networks}. In \bibinfo{booktitle}{\emph{Proceedings of the IEEE Conference on
  Computer Vision and Pattern Recognition}}. \bibinfo{pages}{2891--2900}.
\newblock


\bibitem[\protect\citeauthoryear{Mason}{Mason}{2012}]%
        {mason2012music}
\bibfield{author}{\bibinfo{person}{Paul~H Mason}.}
  \bibinfo{year}{2012}\natexlab{}.
\newblock \showarticletitle{Music, dance and the total art work:
  choreomusicology in theory and practice}.
\newblock \bibinfo{journal}{\emph{Research in dance education}}
  \bibinfo{volume}{13}, \bibinfo{number}{1} (\bibinfo{year}{2012}),
  \bibinfo{pages}{5--24}.
\newblock


\bibitem[\protect\citeauthoryear{McFee, Raffel, Liang, Ellis, McVicar,
  Battenberg, and Nieto}{McFee et~al\mbox{.}}{2015}]%
        {mcfee2015librosa}
\bibfield{author}{\bibinfo{person}{Brian McFee}, \bibinfo{person}{Colin
  Raffel}, \bibinfo{person}{Dawen Liang}, \bibinfo{person}{Daniel~PW Ellis},
  \bibinfo{person}{Matt McVicar}, \bibinfo{person}{Eric Battenberg}, {and}
  \bibinfo{person}{Oriol Nieto}.} \bibinfo{year}{2015}\natexlab{}.
\newblock \showarticletitle{librosa: Audio and music signal analysis in
  python}. In \bibinfo{booktitle}{\emph{Proceedings of the 14th python in
  science conference}}, Vol.~\bibinfo{volume}{8}.
\newblock


\bibitem[\protect\citeauthoryear{Min and Chai}{Min and Chai}{2012}]%
        {min2012motion}
\bibfield{author}{\bibinfo{person}{Jianyuan Min} {and}
  \bibinfo{person}{Jinxiang Chai}.} \bibinfo{year}{2012}\natexlab{}.
\newblock \showarticletitle{Motion graphs++: a compact generative model for
  semantic motion analysis and synthesis}.
\newblock \bibinfo{journal}{\emph{ACM Transactions on Graphics (TOG)}}
  \bibinfo{volume}{31}, \bibinfo{number}{6} (\bibinfo{year}{2012}),
  \bibinfo{pages}{153}.
\newblock


\bibitem[\protect\citeauthoryear{Ofli, Erzin, Yemez, and Tekalp}{Ofli
  et~al\mbox{.}}{2011}]%
        {ofli2011learn2dance}
\bibfield{author}{\bibinfo{person}{Ferda Ofli}, \bibinfo{person}{Engin Erzin},
  \bibinfo{person}{Y{\"u}cel Yemez}, {and} \bibinfo{person}{A~Murat Tekalp}.}
  \bibinfo{year}{2011}\natexlab{}.
\newblock \showarticletitle{Learn2dance: Learning statistical music-to-dance
  mappings for choreography synthesis}.
\newblock \bibinfo{journal}{\emph{IEEE Transactions on Multimedia}}
  \bibinfo{volume}{14}, \bibinfo{number}{3} (\bibinfo{year}{2011}),
  \bibinfo{pages}{747--759}.
\newblock


\bibitem[\protect\citeauthoryear{Peng, Abbeel, Levine, and van~de Panne}{Peng
  et~al\mbox{.}}{2018}]%
        {peng2018deepmimic}
\bibfield{author}{\bibinfo{person}{Xue~Bin Peng}, \bibinfo{person}{Pieter
  Abbeel}, \bibinfo{person}{Sergey Levine}, {and} \bibinfo{person}{Michiel
  van~de Panne}.} \bibinfo{year}{2018}\natexlab{}.
\newblock \showarticletitle{Deepmimic: Example-guided deep reinforcement
  learning of physics-based character skills}.
\newblock \bibinfo{journal}{\emph{ACM Transactions on Graphics (TOG)}}
  \bibinfo{volume}{37}, \bibinfo{number}{4} (\bibinfo{year}{2018}),
  \bibinfo{pages}{143}.
\newblock


\bibitem[\protect\citeauthoryear{Peng, Berseth, Yin, and Van De~Panne}{Peng
  et~al\mbox{.}}{2017}]%
        {peng2017deeploco}
\bibfield{author}{\bibinfo{person}{Xue~Bin Peng}, \bibinfo{person}{Glen
  Berseth}, \bibinfo{person}{KangKang Yin}, {and} \bibinfo{person}{Michiel Van
  De~Panne}.} \bibinfo{year}{2017}\natexlab{}.
\newblock \showarticletitle{Deeploco: Dynamic locomotion skills using
  hierarchical deep reinforcement learning}.
\newblock \bibinfo{journal}{\emph{ACM Transactions on Graphics (TOG)}}
  \bibinfo{volume}{36}, \bibinfo{number}{4} (\bibinfo{year}{2017}),
  \bibinfo{pages}{41}.
\newblock


\bibitem[\protect\citeauthoryear{Safonova and Hodgins}{Safonova and
  Hodgins}{2007}]%
        {safonova2007construction}
\bibfield{author}{\bibinfo{person}{Alla Safonova} {and}
  \bibinfo{person}{Jessica~K Hodgins}.} \bibinfo{year}{2007}\natexlab{}.
\newblock \showarticletitle{Construction and optimal search of interpolated
  motion graphs}.
\newblock \bibinfo{journal}{\emph{ACM Transactions on Graphics (TOG)}}
  \bibinfo{volume}{26}, \bibinfo{number}{3} (\bibinfo{year}{2007}),
  \bibinfo{pages}{106}.
\newblock


\bibitem[\protect\citeauthoryear{Shiratori, Nakazawa, and Ikeuchi}{Shiratori
  et~al\mbox{.}}{2006}]%
        {shiratori2006dancing}
\bibfield{author}{\bibinfo{person}{Takaaki Shiratori}, \bibinfo{person}{Atsushi
  Nakazawa}, {and} \bibinfo{person}{Katsushi Ikeuchi}.}
  \bibinfo{year}{2006}\natexlab{}.
\newblock \showarticletitle{Dancing-to-music character animation}. In
  \bibinfo{booktitle}{\emph{Computer Graphics Forum}},
  Vol.~\bibinfo{volume}{25}. Wiley Online Library, \bibinfo{pages}{449--458}.
\newblock


\bibitem[\protect\citeauthoryear{Tang, Jia, and Mao}{Tang
  et~al\mbox{.}}{2018}]%
        {tang2018dance}
\bibfield{author}{\bibinfo{person}{Taoran Tang}, \bibinfo{person}{Jia Jia},
  {and} \bibinfo{person}{Hanyang Mao}.} \bibinfo{year}{2018}\natexlab{}.
\newblock \showarticletitle{Dance with Melody: An LSTM-autoencoder Approach to
  Music-oriented Dance Synthesis}. In \bibinfo{booktitle}{\emph{2018 ACM
  Multimedia Conference on Multimedia Conference}}. ACM,
  \bibinfo{pages}{1598--1606}.
\newblock


\bibitem[\protect\citeauthoryear{Tieleman and Hinton}{Tieleman and
  Hinton}{2012}]%
        {tieleman2012lecture}
\bibfield{author}{\bibinfo{person}{Tijmen Tieleman} {and}
  \bibinfo{person}{Geoffrey Hinton}.} \bibinfo{year}{2012}\natexlab{}.
\newblock \showarticletitle{Lecture 6.5-rmsprop: Divide the gradient by a
  running average of its recent magnitude}.
\newblock \bibinfo{journal}{\emph{COURSERA: Neural networks for machine
  learning}} \bibinfo{volume}{4}, \bibinfo{number}{2} (\bibinfo{year}{2012}),
  \bibinfo{pages}{26--31}.
\newblock


\bibitem[\protect\citeauthoryear{Wei, Min, and Chai}{Wei et~al\mbox{.}}{2011}]%
        {wei2011physically}
\bibfield{author}{\bibinfo{person}{Xiaolin Wei}, \bibinfo{person}{Jianyuan
  Min}, {and} \bibinfo{person}{Jinxiang Chai}.}
  \bibinfo{year}{2011}\natexlab{}.
\newblock \showarticletitle{Physically valid statistical models for human
  motion generation}.
\newblock \bibinfo{journal}{\emph{ACM Transactions on Graphics (TOG)}}
  \bibinfo{volume}{30}, \bibinfo{number}{3} (\bibinfo{year}{2011}),
  \bibinfo{pages}{1--10}.
\newblock


\bibitem[\protect\citeauthoryear{Xia, Wang, Chai, and Hodgins}{Xia
  et~al\mbox{.}}{2015}]%
        {xia2015realtime}
\bibfield{author}{\bibinfo{person}{Shihong Xia}, \bibinfo{person}{Congyi Wang},
  \bibinfo{person}{Jinxiang Chai}, {and} \bibinfo{person}{Jessica Hodgins}.}
  \bibinfo{year}{2015}\natexlab{}.
\newblock \showarticletitle{Realtime style transfer for unlabeled heterogeneous
  human motion}.
\newblock \bibinfo{journal}{\emph{ACM Transactions on Graphics (TOG)}}
  \bibinfo{volume}{34}, \bibinfo{number}{4} (\bibinfo{year}{2015}),
  \bibinfo{pages}{119}.
\newblock


\bibitem[\protect\citeauthoryear{Yan, Li, Xiong, Yan, and Lin}{Yan
  et~al\mbox{.}}{2019}]%
        {yan2019convolutional}
\bibfield{author}{\bibinfo{person}{Sijie Yan}, \bibinfo{person}{Zhizhong Li},
  \bibinfo{person}{Yuanjun Xiong}, \bibinfo{person}{Huahan Yan}, {and}
  \bibinfo{person}{Dahua Lin}.} \bibinfo{year}{2019}\natexlab{}.
\newblock \showarticletitle{Convolutional sequence generation for
  skeleton-based action synthesis}. In \bibinfo{booktitle}{\emph{Proceedings of
  the IEEE International Conference on Computer Vision}}.
  \bibinfo{pages}{4394--4402}.
\newblock


\bibitem[\protect\citeauthoryear{Zhuang, Wang, Xia, Chai, and Wang}{Zhuang
  et~al\mbox{.}}{2020}]%
        {zhuang2020music2dance}
\bibfield{author}{\bibinfo{person}{Wenlin Zhuang}, \bibinfo{person}{Congyi
  Wang}, \bibinfo{person}{Siyu Xia}, \bibinfo{person}{Jinxiang Chai}, {and}
  \bibinfo{person}{Yangang Wang}.} \bibinfo{year}{2020}\natexlab{}.
\newblock \showarticletitle{Music2Dance: Music-driven Dance Generation using
  WaveNet}.
\newblock \bibinfo{journal}{\emph{arXiv preprint arXiv:2002.03761}}
  (\bibinfo{year}{2020}).
\newblock


\end{thebibliography}
